# Protocolos de Representación de Conocimiento Impreciso e Incierto en una Base de Metaconocimiento Difuso[1]

**Leoncio Jiménez**[2]   **Yolanda Valdés**[3]   **Jorge Vistoso**[3]   **Germain Lacoste**[4]

[2]Departamento de Computación e Informática, Universidad Católica del Maule, Talca – Chile
ljimenez@spock.ucm.cl
[3]Alumnos Seminaristas, Universidad Católica del Maule, Talca – Chile
ysvr76@yahoo.es,j.vistoso@entelchile.net
[4]Ecole Nationale d'Ingenieurs de Tarbes, Tarbes – France
Germain.Lacoste@enit.fr

**RESUMEN**

El presente artículo da a conocer una serie de protocolos para modelar una base de metaconocimiento difuso basada en un modelo conceptual FuzzyEER [19], que permita estructurar el conocimiento de un dominio en una base de datos relacional Oracle 8 para el tratamiento de conocimiento impreciso e incierto usando, por un lado, el modelo GEFRED (GEneralized model for Fuzzy RElational Databases) [13,14], y por otro lado, el sistema FIRST (Fuzzy Interface for Relational SysTems) [8]. En este sentido, el conocimiento es percibido como un objeto, es decir, descrito a través de atributos y valores. En particular nos hemos concentrado en el dominio de competencias (know-how) relativas a la fabricación de cartulinas estucadas, estudiadas en [27]. El proceso que nos interesa en este artículo, es relativo al proceso de conversión de los productos terminados (PILAS o ROLLOS), a partir de las cartulinas fabricadas, dado que en dicho proceso se utilizan atributos clásicos y atributos difusos para caracterizar la calidad de las cartulinas.

**Palabras Clave**: FIRST, Bases de datos relacionales difusas, GEFRED, FuzzyEER.

---

[1] El primer autor es miembro de la Red Iberoamericana de Tecnologías del Software para la década del 2000 (RITOS2).

**1    Introducción**

Una base de datos relacional [9] permite modelar un tipo de datos que, generalmente, se denominan *clásicos*, tales como datos precisos, llamados CRISP (que son escalares simples, por ejemplo: 41), datos desconocidos, no definidos o nulos, todos ellos representados por las palabras reservadas: UNKNOWN, UNDEFINED y NULL, en un entorno Oracle. Sin embargo, estos tipos de datos no son suficientes, para soportar un tipo de datos que manifieste cierta imprecisión e/o incertidumbre, tanto en su representación como en su tratamiento (manipulación). Uno de los primeros estudios matemáticos sobre la representación y la manipulación de datos imprecisos e inciertos es discutida en [4,16]. Estos datos, con imprecisión e incertidumbre, son conocidos por la comunidad internacional en base de datos, con el término *fuzzy*, que ha sido traducido al español como *difuso* o *borroso*, y al francés como *flou*. En todos estos casos su sentido semántico significa *ambiguo* o *vago*, en el sentido del razonamiento humano, que no hay que confundir con el hecho probabilístico de la ocurrencia o no de un evento, según los métodos de la teoría de la probabilidad. Una forma de **representar** este tipo de datos es con el uso la teoría de conjuntos difusos [28], por algunos también llamada teoría de subconjuntos difusos [12] por el echo que el universo del discurso es un conjunto (definido por la teoría de conjuntos), para representar *atributos difusos* según el tipo de referencial que les subyace a los *valores* (ordenado o no ordenado). El supuesto de dicha teoría es que existen conjuntos en los que no está claramente determinado si un elemento pertenece o no al conjunto. A veces, un elemento pertenece al conjunto con cierto grado, llamado grado de pertenencia. Por ejemplo, el conjunto de las personas que son altas es un conjunto difuso, pues no está claro el límite de altura que establece a partir de que medida una persona es alta o no lo es. Ese límite es difuso y, por tanto, el conjunto que delimita también lo será. Un conjunto difuso $A$ sobre un universo de discurso $U$ es un conjunto de pares, dado por: $A = \{m_A(u)/u: u \in U, m_A(u) \in [0,1]\}$, Donde, $m$ es llamada función de pertenencia y $m_A(u)$ es el *grado de pertenencia* del elemento $u$ al conjunto difuso $A$. Este grado oscila entre los extremos 0 y 1, $m_A(u) = 0$, indica que $u$ no pertenece en absoluto al conjunto difuso $A$, $m_A(u) = 1$, indica que $u$ pertenece totalmente al conjunto difuso $A$.

Para **manipular** este tipo de datos se usa la lógica difusa y la teoría de posibilidad [5]. De esta forma han nacido las Bases de Datos Difusas (BDRD), es decir, los sistemas de gestión de bases de datos que tienen la capacidad de trabajar con datos clásicos (precisos, desconocidos, no definidos y nulos) y difusos (imprecisos e inciertos) lo cual puede ser muy ventajoso, ya que el tipo de consultas que puede hacerse a una BDRD puede a semejarse al lenguaje natural. Por ejemplo, en el entorno de calidad del papel, podemos considerar la siguiente consulta "*Seleccione los ROLLOS de cartulina almacenados en la bodega que están sucios y húmedos*". Es claro, que estos criterios de búsqueda pueden ser considerados como ambiguos o vagos, ya que lo que es sucio para una persona, no lo puede ser para otra. En nuestro escenario de estudio, hemos considerado los parámetros altura y diámetro de los ROLLOS como difusos, solamente para ilustrar nuestra propuesta, pero en la realidad (formulario de pedidos) no lo son. Lo mismo, los atributos ancho, largo y altura de las PILAS. Por ejemplo, si consideramos el atributo *diámetro* de un ROLLO, en una base de datos relacional clásica sólo se aceptarán valores exactos, por ejemplo: 400 cm. de diámetro. Ahora, que sucede si 400 cm. es la norma especificada para el ROLLO, y se tienen almacenados otros ROLLOS en la bodega, cuyos diámetros están por encima o por debajo de lo especificado por la norma, el problema parece evidente: ¿Cómo clasificar esos ROLLOS en el sistema?. Es en este tipo de situación que las BDRD se vuelven interesantes, ya que es posible definir un conjunto difuso {"debajo-de-la-norma", "en-la-norma", "arriba-de-la-norma"}, en que la pertenencia de un valor, por ejemplo: 402 cm. de diámetro, a los subconjuntos "debajo-de-la-norma" y "en-la-norma", pueda ser incierta. Es decir, con un *grado de pertenencia*.

En este tipo de situaciones es donde se producen problemas a la hora de definir la especificación de un modelo conceptual que permita la captura y la representación de datos (clásicos y difusos). Además, los modelos conceptuales propuestos en la literatura [3,6] no permiten "dialogar" directamente con el modelo lógico, siendo necesario una transformación de los datos dependiendo de la plataforma de implementación.

En este sentido, en lo que respecta al modelo lógico de una base de datos relacional difusa, en [18] se presenta un entorno para consultas flexibles, es decir, que presentan atributos difusos en la consulta. Sin embargo, en dicho trabajo no es estudiado el modelo conceptual. Lo mismo ocurre en los trabajos de [1,2,8]. En efecto, en ellos se muestran dos lenguajes de consultas difusas a partir de una extensión del lenguaje SQL (Structured Query Language) a una base de datos relacional. Ellos son: SQLf y FSQL. Siendo, este último el único que incorpora un módulo "difuso" a la plataforma Oracle, para poder almacenar y consultar información difusa, pero en ambos casos, no se aborda el modelo conceptual. De igual forma, en [13,14,15] se discuten los aspectos del diseño lógico pero nada se propone para el diseño conceptual de una base de datos relacional difusa. Recientemente, en [21,22] se propone una extensión del modelo EER (Enhanced Entity Relationship), que permite una *notación* para los atributos difusos, que puedan ser soportados en el diseño lógico en FSQL. Además, en [19] se muestra una herramienta CASE, llamada FuzzyCASE que soporta dicha *notación*. En lo que respecta, a la utilización de FSQL para la gestión, éste ha sido utilizado en el área del datamaning [7], para la gestión de datos difusos al caso de una agencia inmobiliaria. En donde, la *notación* [23] propuesta para el modelo conceptual ha sido aplicada [20] a ese mismo caso de estudio para representar los atributos difusos de la agencia inmobiliaria, mientras que en [25,26] se ofrece una notación para el modelado de datos difusos en UML (diagrama de caso de uso). Por último, en [10,11] se muestra una aplicación de dicha *notación* al modelado conceptual de un sistema basado en conocimiento usando la metodología CommonKADS. El interés de este trabajo es mostrar un escenario para la gestión del conocimiento en una base de datos relacional difusa. Para ello hemos simplificado un domino de conocimiento que es discutido en [27] para mostrar las características que deben tener los atributos clásicos y atributos difusos, en un diseño en tablas (modelo lógico), para su respectiva implementación en Oracle 8.

El trabajo esta organizado en tres apartados. En el primero, se muestra la representación de atributos difusos que hemos adoptado para el modelo conceptual, y que se basa en nuestros trabajos previas. En el segundo apartado, se describe el modelo FuzzyEER que corresponde al esquema conceptual de la bases de datos relacional difusa. En el tercer apartado, se describe el modelo lógico.

## 2    Representación de atributos difusos

En [21,23] se propone una extensión del modelo EER (Enhanced Entity Relationship), llamado FuzzyEER, que permite modelar atributos difusos y su almacenamiento lo encontramos en [8]. Los atributos difusos pueden ser de tres tipos:

- **Tipo 1**: Estos atributos son tradicionales o crisp, pero pueden efectuar consultas flexibles utilizándolos en las condiciones difusas. Además sobre su dominio podemos definir etiquetas lingüísticas, asociadas a algún valor difuso concreto.
- **Tipo 2**: Estos atributos admiten tanto datos crisp como difusos, en forma de distribución de posibilidad sobre un dominio ordenado. Ellos permiten almacenar y usar todo los tipos de constantes difusas.
- **Tipo 3**: Estos atributos admiten valores que son definidos sobre un dominio no ordenado. Este tipo de atributo es usado para almacenar escalares con una relación de similitud definida sobre ellas.

También se pueden almacenar los valores siguientes: el valor UNKNOWN si no se sabe nada sobre el valor de un determinado atributo. El valor UNDEFINED se utiliza si un determinado valor no es aplicable o carece de sentido. El valor NULL será utilizado si no se sabe nada sobre ese atributo, eso es, si ignoramos si es o no aplicable.

## 3    Modelo FuzzyEER

Consideramos la siguiente simplificación del dominio de conocimiento, estudiado en [27].

Los atributos de la cartulina se refieren a las cualidades que deben tener los productos terminados (PILAS o ROLLOS). Estas cualidades no son medidas con instrumentos físicos, ya que, sólo se pueden apreciar con los sentidos humanos, principalmente la vista y el tacto. Estas cualidades son las que le confieren las aptitudes para satisfacer las necesidades del uso de las cartulinas implícitas. Estas cualidades están preestablecidas y son parte del know-how de la empresa. La tabla 1 muestra algunos de los atributos que están implícitos en la cartulina.

| Atributos de una PILA | Atributos de un ROLLO |
|---|---|
| Pliegos parejos y planos. | Bobinado parejo. |
| Pliegos planos. | Limpio. |
| Tarima seca y con dimensiones correctas. | Empalmes correctos (bien pegados e identificados). |
| Cantidad exacta. | Corte limpio (no pelusiento). |
| Formato especificado (ancho, largo, altura). | Formato especificado (diámetro, altura). |
| Identificación correcta. | Identificación correcta. |

**Tabla 1:** Características de la PILA y ROLLO.

La figura 1 muestra el modelo FuzzyEER (esquema conceptual) para la característica *formato especificado* de la tabla 1. En este esquema conceptual se distinguen dos subclases (o subtipos) disjuntas de la entidad cartulinas, ya que, por ejemplo una cartulina que es un rollo no puede ser una pila a la vez. Para cada una de las entidades del modelo FuzzyEER de la figura 1, la representación de sus atributos difusos es como sigue:

**Entidad Cartulinas:** Definida por el esquema {Código cartulina, Tono cara, Tono reverso, Impresión}. En que *Código cartulina* corresponde a un dato CRISP. Mientras que *Tono cara* y *Tono reverso* son atributos de Tipo 3 por ser de dominios subyacentes no ordenados con las etiquetas escalares {blanco, amarillo, café, manila}. En efecto, es posible definir una relación de semejaza entre ellos. Estos atributos hacen referencia al color por ambos lados de la cartulina. El atributo *Impresión* representa el tipo de impresión recomendada para las cartulinas. En este caso se trata de un atributo Tipo 1 definido para el conjunto de etiquetas {huecograbado, offset}.

**Entidad Pilas:** Definida por el esquema {Código cartulina, Código pila, Largo, Ancho, Altura, Peso, Estado}. En que *Código cartulina* y *Código pila* corresponden a dos datos CRISP. Mientras que L*argo* es un atributo de Tipo 2 definido por las etiquetas {corto, largo, muy largo}; A*ncho* es un atributo de Tipo 2 definido por las etiquetas {angosto, ancho, muy ancho}; *Altura* es un atributo de Tipo 2 definido por las etiquetas {alta, muy alta, baja, muy baja}; y *Peso* es un atributo de Tipo 2 definido por las etiquetas {bajo, muy bajo, sobre, muy sobre}, ya que en todos esos casos se trata de atributos cuyo dominio es ordenado. El atributo *Estado* es considerado de Tipo 3 por ser de dominio subyacente no ordenado con las etiquetas escalares {golpeado, mojado, orilla picada, englobado, sucio, picaduras, rayas en la superficie}. En efecto, es posible definir una relación de semejaza entre ellos.

**Entidad Rollos:** Definida por el esquema {Código cartulina, Código rollo, Diámetro, Altura, Peso, Estado}. En que *Código cartulina* y *Código rollo* corresponden a dos datos CRISP. Mientras que *Diámetro* es un atributo de Tipo 2 definido por las etiquetas {rango mínimo, normal, rango máximo}; *Altura* es un atributo de Tipo 2 definido por las etiquetas {baja, mediana, alta}; y *Peso* es un atributo de Tipo 2 definido por las etiquetas {bajo, optimo, sobre}. El atributo *Estado* es considerado de Tipo 3 por ser de dominio subyacente no ordenado con las etiquetas escalares {englobado, deslaminado, húmedo, sucio, rayas, curvas, empalme defectuoso, orilla crespa, disparejo}. En efecto, es posible definir una relación de semejaza entre ellas.

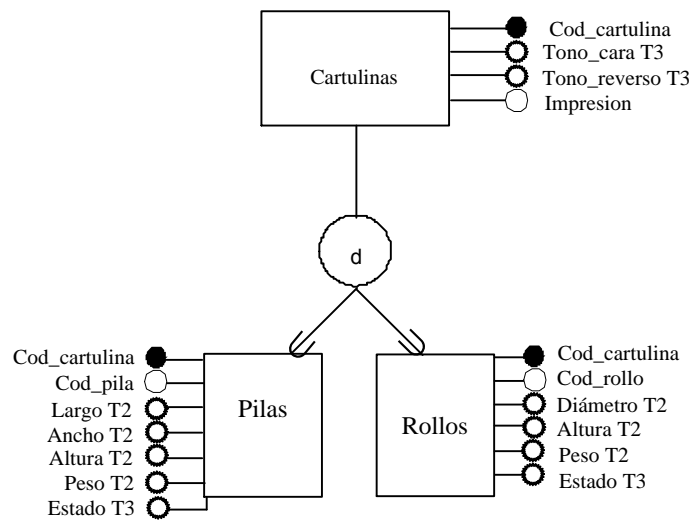

**Figura 1:** Modelo FuzzyEER.

## 4   Modelo lógico

A partir del modelo conceptual, definido en el párrafo 3, se construye el modelo lógico de la base de datos relacional difusa. Por una parte, se deben representar los atributos difusos Tipo 1, Tipo 2 y Tipo 3, en la base de metaconocimiento difuso (FMB, Fuzzy Metaknowledge Base), por medio de un protocolo de representación para cada uno esos atributos. Esto se consigue con el sistema FIRST (Fuzzy Interface for Relational SysTems) propuesto por [12,14], y mejorado por [8], gracias a la integración del módulo FSQL (Fuzzy Structured Query Language) en FIRST para generar consultas flexibles que puedan utilizar cuantificadores difusos. Para ello, el FSQL se compone de una arquitectura Cliente – Servidor, que permite la conexión entre la base de datos Oracle y la base de metaconocimiento difuso. Por otra parte, se debe almacenar dichos atributos difusos en la base de metaconocimiento difuso, por medio de un script en lenguaje PL/SQL, es decir que permite el cargado de las tablas. Un estudio más detallado de la FIRST y FSQL, se encuentra en [8]. A continuación detallamos la representación y el almacenamiento de los atributos difusos de la figura 1 en la base de metaconocimiento difuso.

### 4.1.   Representación de la FMB (Fuzzy Metaknowledge Base)

La FMB es un módulo del sistema FIRST que se utiliza como un "diccionario" o "catalogo" para describir los datos de los atributos difusos Tipo 1, Tipo 2 y Tipo 3 del modelo conceptual en el SGBDR. Por lo tanto, dicho módulo contiene el protocolo de transformación para los tres atributos difusos definidos en el párrafo 3. La organización de la FMB es mediante el uso de tablas o relaciones. Para la claridad del artículo hemos modificado la descripción de las tablas. La versión original puede ser consultada en [8]. A continuación detallamos cada uno de los protocolos de representación de los datos de los atributos difusos Tipos 1, 2 y 3 de la FMB.

#### 4.1.1.   Protocolo de Representación para el Atributo Difuso Tipo 1

Para este tipo de atributo no existe una tabla de conversión, ya que los datos reciben una representación igual que los datos precisos. Sin embargo, en el diccionario FMB se almacena información relativa a las etiquetas de los atributos difusos Tipo 1, también recogerá información acerca de la naturaleza de estos atributos. Por consiguiente, son atributos clásicos que admiten el tratamiento difuso y por tanto podremos efectuar consultas (flexibles) sobre ellos, aunque no permita almacenar valores difusos.

#### 4.1.2.   Protocolo de Representación para el Atributo Difuso Tipo 2

La tabla 2 muestra el protocolo para almacenar este tipo de datos. La primera columna muestra la familia de datos que es posible almacenar con este atributo. La segunda columna muestra el identificador asociado a cada atributo. Por ejemplo, el dato APROXIMADAMENTE tiene asociado el identificador 6. La tercera columna, se subdivide en cuatro columnas, cada una de ellas almacena la descripción de las variables que definen el dato, y depende, bien entendido, del tipo de dato. Lo que significa, para el ejemplo anterior, que necesitamos 4 variables para almacenar un atributo difuso (#d en FSQL). La primera columna almacena el valor (d), la segunda columna su limite izquierdo (d-margen), la tercera columna su limite derecho (d+margen), y la cuarta columna el margen. Los valores NULL que aparecen en las otras familias de datos Tipo 2 de la tabla 2, tienen el significado de valor "no-aplicable" en el sistema SGBDR anfitrión.

| Familia de los Datos Tipo 2 | Protocolo de los Datos Tipo 2 | | | | |
|---|---|---|---|---|---|
| | Id | V1 | V2 | V3 | V4 |
| UNKNOWN | 0 | NULL | NULL | NULL | NULL |
| UNDEFINED | 1 | NULL | NULL | NULL | NULL |
| NULL | 2 | NULL | NULL | NULL | NULL |
| CRISP | 3 | d | NULL | NULL | NULL |
| LABEL | 4 | FUZZY_ID | NULL | NULL | NULL |
| INTERVALO (n,m) | 5 | n | NULL | NULL | m |
| APROXIMADAMENTE (d) | 6 | d | d-margen | d+margen | margen |
| TRAPECIO | 7 | $\alpha$ | $\beta-\alpha$ | $\gamma-\delta$ | $\delta$ |

**Tabla 2:** Protocolos de Representación en la FMB de los Datos para el Atributo Difuso Tipo 2.

### 4.1.3. Protocolo de Representación para el Atributo Difuso Tipo 3

La tabla 3 muestra el protocolo para almacenar este tipo de datos. La primera columna muestra la familia de datos que es posible almacenar con un atributo Tipo 3. La segunda columna muestra el identificador asociado a cada atributo. Por ejemplo, el dato DISTRIBUCION de POSIBILIDAD, tiene asociado el identificador 4. La tercera columna, se subdivide en *n* columnas de *n* parejas, con *n*≥1, del tipo (*valor de posibilidad, etiqueta*), (FP1, F1),..., (FPn, Fn), donde es posible almacenar los valores de la distribución de posibilidad. Estos valores se encuentran en el intervalo [0, 1]. Por otra parte, un dato de tipo SIMPLE sólo se usa la primera pareja. Sin embargo, se puede usar menos de *n* parejas dejando el resto de campos a NULL. En la FMB se almacenan las *etiquetas*, su relación de semejanza y el valor de *n*.

| Familia de los Datos Tipo 3 | Protocolo de los Datos Tipo 3 | | | | | |
|---|---|---|---|---|---|---|
| | FT | FP1 | F1 | ... | $FP_n$ | $F_n$ |
| UNKNOWN | 0 | NULL | NULL | ... | NULL | NULL |
| INDEFINED | 1 | NULL | NULL | ... | NULL | NULL |
| NULL | 2 | NULL | NULL | ... | NULL | NULL |
| SIMPLE | 3 | p | d | ... | NULL | NULL |
| DISTRIBUCION POSIBILIDAD | 4 | $p_1$ | $d_1$ | ... | $p_n$ | $d_n$ |

**Tabla 3:** Protocolos de Representación en la FMB de los Datos para el Atributo Difuso Tipo 3.

### 4.2. Almacenamiento de la FMB (Fuzzy Metaknowledge Base)

La figura 2 define el diseño lógico de la FIRST para cada etiqueta lingüística de las PILAS o ROLLOS, según lo definido en el modelo FuzzyEER de la figura 1.

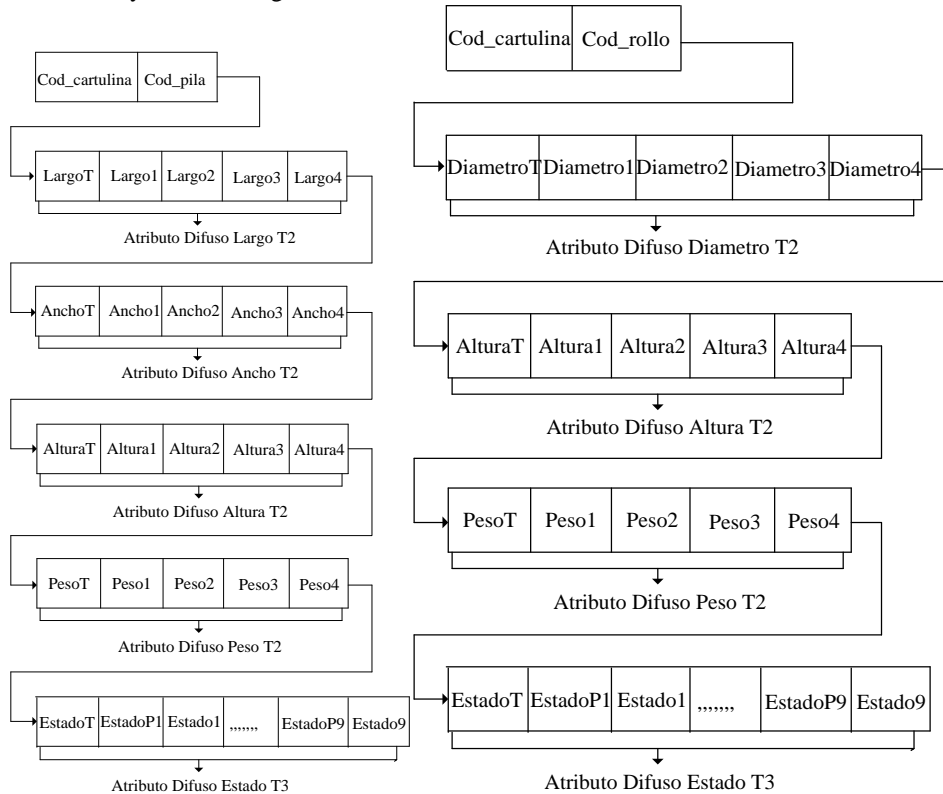

**Figura 2:** Diseño lógico de la FIRST para los atributos difusos del Modelo FuzzyEER (figura 1) para las PILAS y ROLLOS.

Según el diseño lógico de la FIRST (figura 2), las etiquetas lingüísticas deben ser almacenadas en tablas de la FMB. En este caso la FMB está compuesta de las siguientes tablas: FUZZY_COL_LIST (FCL), FUZZY_OBJECT_LIST (FOL), FUZZY_LABEL_DEF (FLD), y FUZZY_NEARNESS_DEF (FND). En que el poblado de las tablas se realiza según el tipo de atributo difuso. La figura 3a) muestra, con la ayuda de un autómata a estado finito, la estrategia de almacenamiento de los atributos difusos de Tipo 2, mientras que la figura 3b) indica la estrategia a seguir para los atributos de Tipo 3.

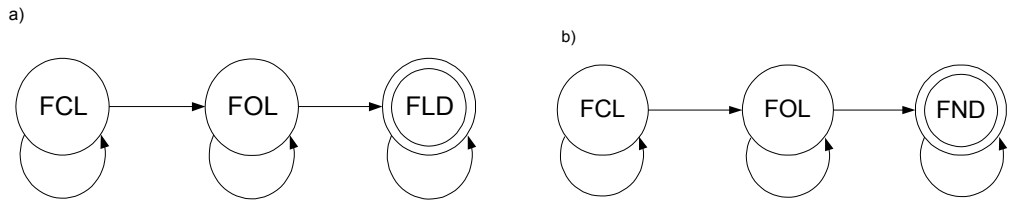

**Figura 3:** Autómata finito para: a) tipo atributo difuso Tipo 2 y b) atributo difuso Tipo 3.

A continuación se detalla la composición de cada una de esas tablas y, un ejemplo de script en lenguaje PL/SQL, que permite el llenado de la FMB.

### 4.2.1. Tabla FUZZY_COL_LIST (FCL)

La tabla FCL contiene la descripción de todos los atributos de la base de datos, que son susceptibles de tratamiento difuso, es decir de todas las relaciones. Esta tabla esta estructurada como sigue. Las columnas OBJ# y COL# permiten almacenar el nombre de la entidad y sus atributos. La columna F_TYPE permite almacenar el tipo de atributo difuso identificado en COL#. Mientras que la columna LEN permite almacenar la longitud máxima de una distribución de posibilidad en atributos Tipo 3. La figura 4 muestra la tabla FCL para el caso de estudio.

| OBJ# | COL# | F_TYPE | LEN |
|---|---|---|---|
| Pilas | Largo | 2 | 1 |
| Pilas | Ancho | 2 | 1 |
| Pilas | Altura | 2 | 1 |
| Pilas | Peso | 2 | 1 |
| Pilas | Estado | 3 | 9 |
| Rollos | Diametro | 2 | 1 |
| Rollos | Altura | 2 | 1 |
| Rollos | Peso | 2 | 1 |
| Rollos | Estado | 3 | 9 |

**Figura 4:** Tabla FCL.

La figura 5 muestra un ejemplo del script en lenguaje PL/SQL que es utilizado en el sistema FIRST para cargar la tabla FCL en la base de datos difusa.

```
INSERT into FCL values (t_PILAS,c_PLARGO,2,1,USER||'.Pilas.largo');
INSERT into FCL values (t_PILAS,c_PANCHO,2,1,USER||'.Pilas.ancho');
INSERT into FCL values (t_PILAS,c_PPESO,2,1,USER||'.Pilas.Peso');
INSERT into FCL values (t_PILAS,c_PESTADO,3,9,USER||'.Pilas.Estado');
```

**Figura 5:** Ejemplo de Script PL/SQL para la tabla FCL.

El script de la figura 5, permite ingresar los atributos difusos. En la primera columna de la tabla FCL (figura 4), que corresponde al campo OBJ#, se almacena la entidad *t_PILAS*, mientras que en la columna COL# se ingresa el atributo de dicha entidad. Para el ejemplo en cuestión de la figura 5, almacenamos: *c_PLARGO*, *c_PANCHO*, *c_PPESO* y *c_PESTADO*. En la columna F_TYPE se almacenan los valores 2 y 3, que corresponden al tipo de datos del atributo, por ejemplo el atributo Largo de la entidad Pila es de Tipo 2. En la columna LEN se almacena el número de parejas de un atributo. Por ejemplo, en caso del atributo Estado de la entidad Pila, se almacena el valor 9, porque corresponde al número de relaciones de similitud entre los datos del atributo, en este caso corresponde al conjunto {Englobado, Deslaminado, Humedad, Sucio, Rayas, Curvas, Empalme_defectuoso, Orilla_crespa, Disparejo}. En los atributos de tipo 2 se debe almacenar el valor 1 como mínimo. En el script se utiliza la palabra reservada USER para identificar a la tupla que se ingresa, por ejemplo, el Largo de la Pila queda identificado por '.Pilas.largo'.

### 4.2.2. Tabla FUZZY_OBJECT_LIST (FOL)

La tabla FOL contiene la descripción de todas las etiquetas definidas para cada uno de las relaciones definidas en la tabla FCL. En lo que se refiere a su organización, la clave primaria de acceso a esta tabla esta dada por las columnas OBJ# y COL#. La columna FUZZY_ID permiten almacenar el identificador (0,1,2,...) del valor del atributo que es especificado en la columna FUZZY_NAME. Mientras que en FUZZY_TYPE está el tipo de dato. Por ejemplo, para las etiquetas trapezoidales se almacena un 0, en cambio para una relación de similitud el valor a almacenar es 1. En general, esto se encuentra definido en la tabla FLD. La figura 6 muestra la tabla FOL para el caso de estudio.

| OBJ# | COL# | FUZZY_ID | FUZZY_NAME | FUZZY_TYPE |
|---|---|---|---|---|
| Rollos | Diametro | 0 | 'Rango_min' | 0 |
| Rollos | Diametro | 1 | 'Normal' | 0 |
| Rollos | Diametro | 2 | 'Rango_max' | 0 |
| Rollos | Altura | 0 | 'Baja' | 0 |
| Rollos | Altura | 1 | 'Mediana' | 0 |
| Rollos | Altura | 2 | 'Alta' | 0 |
| Rollos | Peso | 0 | 'Bajo' | 0 |
| Rollos | Peso | 1 | 'Optimo' | 0 |
| Rollos | Peso | 2 | 'Sobre' | 0 |
| Rollos | Estado | 0 | 'Englobado' | 1 |
| Rollos | Estado | 1 | 'Deslaminado' | 1 |
| Rollos | Estado | 2 | 'Humedo' | 1 |
| Rollos | Estado | 3 | 'Sucio' | 1 |
| Rollos | Estado | 4 | 'Rayas' | 1 |
| Rollos | Estado | 5 | 'Curvas' | 1 |
| Rollos | Estado | 6 | 'Empalme_defectuoso | 1 |
| Rollos | Estado | 7 | 'Orilla_crespa' | 1 |
| Rollos | Estado | 8 | 'Disparejo' | 1 |

**Figura 6:** Tabla FOL.

La figura 7 muestra un ejemplo del script en lenguaje PL/SQL que es utilizado en el sistema FIRST para cargar la tabla FOL en la base de datos difusa.

```
INSERT into FOL values(t_ROLLOS,c_RDIAMETRO,0,'RANGO_MIN',0);
INSERT into FOL values(t_ROLLOS,c_RDIAMETRO,1,'NORMA',0);
INSERT into FOL values(t_ROLLOS,c_RDIAMETRO,2,'RANGO_MAX',0);
INSERT into FOL values(t_ROLLOS,c_RESTADO,7,'ORILLA_CRESPA',1);
INSERT into FOL values(t_ROLLOS,c_RESTADO,8,'DISPAREJO',1);
```

**Figura 7:** Ejemplo de Script PL/SQL para la tabla FOL.

El script de la figura 7, permite ingresar las etiquetas lingüísticas de los atributos difusos. En la primera columna de la tabla FOL (figura 6), que corresponde al campo OBJ#, se almacena la entidad *t_ROLLOS*, mientras que en la columna COL# se ingresa el atributo de dicha entidad. Para el ejemplo en cuestión de la figura 6, almacenamos:

*c_RDIAMETRO* y *c_RESTADO*. En la columna FUZZY_ID se almacenan los valores 0, 1 y 2, que corresponden al identificador para cada una de las etiquetas del atributo, las cuales son almacenadas en la columna FUZZY_NAME. En el ejemplo, para el atributo diámetro se tienen las etiquetas {Rango_min, Norma, Rango_max}. Finalmente, en la columna FUZZY_TYPE se almacena un 0, si el atributo es de Tipo 2, o un 1 si el atributo es Tipo 3.

### 4.2.3. Tabla FUZZY_LABEL_DEF (FLD)

La tabla FLD contiene la descripción de todas las etiquetas definidas sobre un atributo Tipo 2, para cada uno de los atributos Tipo 2 de la tabla FOL. Es decir, esta tabla contiene los puntos que determinan la distribución de posibilidad trapezoidal para los FUZZY_TYPE=0 de la tabla FOL. Los cuales, son ingresados en las columnas ALFA, BETA, GAMMA, DELTA. Por otra parte, la clave primaria de acceso a esta tabla esta dada por las columnas OBJ#, COL# y FUZZY_ID de la tabla FOL. La figura 8 muestra la tabla FLD para el caso de estudio.

| OBJ# | COL# | FUZZY_ID | ALFA | BETA | GAMMA | DELTA |
|---|---|---|---|---|---|---|
| Rollos | Diametro | 0 | 50 | 70 | 100 | 130 |
| Rollos | Diametro | 1 | 100 | 150 | 170 | 220 |
| Rollos | Diametro | 2 | 190 | 220 | 250 | 300 |
| Rollos | Altura | 0 | 3 | 4 | 5 | 7 |
| Rollos | Altura | 1 | 5 | 8 | 10 | 11 |
| Rollos | Altura | 2 | 10 | 12 | 15 | 17 |
| Rollos | Altura | 3 | 14 | 16 | 17 | 19 |
| Rollos | Peso | 0 | 15 | 20 | 35 | 40 |
| Rollos | Peso | 1 | 30 | 45 | 65 | 75 |
| Rollos | Peso | 2 | 70 | 85 | 95 | 100 |

**Figura 8:** Tabla FLD.

La figura 9 muestra un ejemplo del script en lenguaje PL/SQL que es utilizado en el sistema FIRST para cargar la tabla FLD en la base de datos difusa.

```
INSERT into FLD values(t_ROLLOS,c_RDIAMETRO,0,50,70,100,130);
INSERT into FLD values(t_ROLLOS,c_RDIAMETRO,1,100,150,170,220);
INSERT into FLD values(t_ROLLOS,c_RDIAMETRO,2,190,220,250,300);
```

**Figura 9:** Ejemplo de Script PL/SQL para la tabla FLD.

El script de la figura 9, permite ingresar los valores de las etiquetas lingüísticas. En la primera y segunda columna de la tabla FLD (figura 8), que corresponden a los campos OBJ# y COL#, se almacenan la entidad y su atributo, respectivamente, en el ejemplo *t_ROLLOS* y *c_RDIAMETRO*. En la columna FUZZY_ID se almacenan los identificadores de la tabla FOL (figura 6). En las columnas ALFA, BETA, GAMMA y DELTA se almacenan los valores que representan al trapecio de la etiqueta, por ejemplo, en el diámetro del rollo que posee como identificador el valor 0, y que corresponde al FUZZY_NAME 'Rango_min' de la tabla FOL (figura 6), tiene como valores de la etiqueta 50, 70, 100, 130.

#### 4.2.4. Tabla FUZZY_NEARNESS_DEF (FND)

La tabla FND contiene la descripción de todas las medidas de semejanza entre las etiquetas definidas sobre un atributo Tipo 3 de la tabla FOL. Es decir, esta tabla contiene los grados de similitud por cada pareja de etiquetas distintas para los cuales FUZZY_TYPE=1 en la tabla FOL. La clave primaria de acceso a esta tabla esta dada por las columnas OBJ#, COL#, FUZZY_ID1 y FUZZY_ID2 de la tabla FOL. La figura 10 muestra la tabla FND para el caso de estudio.

| OBJ#   | COL#   | FUZZY_ID1 | FUZZY_ID2 | DEGREE |
|--------|--------|-----------|-----------|--------|
| Rollos | Estado | 0         | 2         | 0      |
| Rollos | Estado | 0         | 3         | 0      |
| Rollos | Estado | 0         | 4         | 0      |
| Rollos | Estado | 0         | 5         | 0.3    |
| Rollos | Estado | 0         | 6         | 0.5    |
| Rollos | Estado | 0         | 7         | 0.6    |
| Rollos | Estado | 0         | 8         | 0      |
| Rollos | Estado | 1         | 2         | 0      |
| Rollos | Estado | 1         | 3         | 0      |
| Rollos | Estado | 1         | 4         | 0      |
| Rollos | Estado | 1         | 5         | 0      |
| Rollos | Estado | 1         | 6         | 0.8    |
| Rollos | Estado | 1         | 7         | 0      |
| Rollos | Estado | 1         | 8         | 0.1    |

**Figura 10:** Tabla FND.

La figura 11 muestra un ejemplo del script en lenguaje PL/SQL que es utilizado en el sistema FIRST para cargar la tabla FND en la base de datos difusa.

```
INSERT into FND values(t_ROLLOS,c_RESTADO,0,2,0);
INSERT into FND values(t_ROLLOS,c_RESTADO,0,3,0);
INSERT into FND values(t_ROLLOS,c_RESTADO,0,4,0);
INSERT into FND values(t_ROLLOS,c_RESTADO,0,5,.3);
INSERT into FND values(t_ROLLOS,c_RESTADO,0,6,.5);
```

**Figura 11:** Ejemplo de Script PL/SQL para la tabla FND.

El script de la figura 11, permite ingresar los valores de la relación de similitud para atributos de Tipo 3. En la primera y segunda columna de la tabla FND (figura 10), que corresponden a los campos OBJ# y COL#, se almacenan la entidad y su atributo, respectivamente, en el ejemplo *t_ROLLOS* y *c_RESTADO*. Las columnas FUZZY_ID1 y FUZZY_ID2 son los identificadores de las etiquetas definidas en la tabla FOL (figura 6). Así, por ejemplo, en el cuarto INSERT de la figura 11, se tiene 0,5,.3, que significa que la etiqueta 0 para "Englobado" y la etiqueta 5 para "Curvas", tienen una relación de similitud 0.3. Dicho valor se almacena en la columna DEGREE.

## 5. Conclusiones

En este artículo se ha presentado una serie de protocolos para modelar una base de metaconocimiento difuso, para ello se ha utilizado un ejemplo simplificado del modelo FuzzyEER estudiado [27], luego en el diseño lógico de la base de datos relacional difusa, se ha establecido una estrategia para el poblado de las tablas de la FMB. Este trabajo constituye una evolución en el saber hacer del grupo de bases de datos de nuestro departamento en lo referente al manejo de la FIRST y la programación en lenguaje PL/SQL de la base de metaconocimiento difuso, iniciados en [17]. Sin dejar de mencionar, que este trabajo extiende, por una parte el estudio [21,23] del modelado conceptual de datos al diseño lógico, incorporando en el diseño lógico los atributos difusos Tipo 3, que no se encuentran especificados [8], y por otra parte, lo desarrollado en [24] relativo a la gestión de conocimiento y lo desarrollado en [10,11] que muestra una herramienta de la inteligencia artificial para ser utilizada en el modelado de datos imprecisos e inciertos en un dominio industrial.

Como trabajo futuro, se contempla la implantación de un sistema de control de calidad para la empresa en estudio, que permita la gestión de conocimiento impreciso e incierto.

## 6. Referencias

**Agrecedimientos**